\documentclass[12pt]{amsart}

\usepackage[dvips]{epsfig}
\usepackage{amsmath}
\usepackage{amssymb}
\usepackage{amsfonts}
\usepackage{amsbsy}
\usepackage{amsgen}
\usepackage{amscd}
\usepackage{amsopn}
\usepackage{amstext}
\usepackage{amsxtra}
\usepackage{enumerate}
\usepackage{dsfont}
\usepackage{lineno}

\newtheorem{theorem}{Theorem}[section]
\newtheorem{proposition}[theorem]{Proposition}

\newtheorem{corollary}[theorem]{Corollary}
\theoremstyle{definition}

\newtheorem{conjecture}[theorem]{Conjecture}
\newtheorem{question}[theorem]{Question}

\newcommand {\E} {\mathbb{E}}

\newcommand {\R} {\mathbb{R}}
\newcommand {\Z} {\mathbb{Z}}

\begin{document}


\author{P\"ar Kurlberg}
\address{Department of Mathematics, KTH Royal Institute of Technology,
SE-100 44 Stockholm, Sweden}

\author{Igor Wigman}
\address{Department of Mathematics, King's College London, UK}


\title{Non-universality of the Nazarov-Sodin constant}

\date{June 29, 2017}

\begin{abstract}
We prove that the {\em Nazarov-Sodin
    constant}, which up to a natural scaling gives the leading order
  growth for the expected number of nodal components
 of a random Gaussian field, genuinely depends on the field. We then
  infer the same for ``arithmetic random waves'', i.e. random toral
  Laplace eigenfunctions.

%
\end{abstract}

\maketitle

\section{The Nazarov-Sodin constant}

Let $m\ge 2$, and $$f:\R^{m}\rightarrow\R$$ be a {\em stationary}
centred Gaussian random field, and $r_{f}:\R^{m}\rightarrow \R$ its
covariance function defined as $$r_{f}(x)= \E[f(y)f(y+x)].$$ Given
such an $f$, let $\rho = \rho_{f}$ denote its spectral measure,
i.e. the Fourier transform of $r_{f}$ (assumed to be a probability
measure); note that prescribing $\rho$ defines $f$ uniquely.  We
further assume that a.s. $f$ is sufficiently smooth, and that the
distribution of $\nabla f(x)$ is non-degenerate.

Let $N(f;R)$ be the number of connected components of
$f^{-1}(0)$ in $B_{0}(R)$ (the radius-$R$ ball centred at $0$),
usually referred to as the {\em nodal
components} of $f$; $N(f;R)$ is a random variable. Nazarov and
Sodin \cite[Theorem $1$]{So 2012} proved that under the above
conditions the expected number of nodal components of $f$ is
\begin{equation}
\label{eq:EN(f;R)=cNSR^m+o(Rm)}
\E[N(f;R)] = c_{NS}(\rho_{f})R^{m}+o(R^{m}),
\end{equation}
where $c_{NS}(\rho_{f}) \ge 0$ is referred to as the
Nazarov-Sodin constant of $f$ (we will consider
$c_{NS}$ as a function of the spectral density of $\rho_{f}$ rather than
of $f$).

For $m=2$, $\rho=\rho_{\mathcal{S}^{1}}$ the
uniform measure on the unit circle $\mathcal{S}^{1}\subseteq \R^{2}$
(i.e. $d\rho=\frac{d\theta}{2\pi}$ on $\mathcal{S}^{1}$ vanishing
outside the circle) the corresponding random field $f_{\text{RWM}}$ is
known as {\em random monochromatic waves}; Berry ~\cite{Berry 1977}
suggested that $f_{\text{RWM}}$ may serve as a {\em universal model}
to Laplace eigenfunctions on generic surfaces in the high energy limit
--- the {\em Random Wave Model}. The corresponding {\em universal}
Nazarov-Sodin constant $c_{\text{RWM}}$ is known to be strictly
positive, and in \cite{BS} its value was predicted using a certain
percolation model. However, recent numerics by Nastacescu, as well as
by Konrad, show a small deviation from these
predictions.

\vspace{3mm}

Let $(\mathcal{M}^{m},g)$ be a smooth manifold. Here the
restriction of a fixed random field $f:\mathcal{M}\rightarrow \R$ to
growing domains, as was considered on the Euclidean space, makes no
sense. Instead we consider a sequence of random fields $\{f_{L}
\}_{L\in \mathcal{L}}$ (for $L$ lying in some discrete subset
$\mathcal{L}\subseteq\R$), and the total number $N(f_{L})$ of nodal
components of $f_{L}$ on $M$.  Here we may define a scaled covariance
function of $f_{L}$ around a fixed point $x\in\mathcal{M}$ on its
tangent space $T_{x}(\mathcal{M})\cong \R^{m}$ via the exponential map
at $x$, and assume that for a.e. $x\in\mathcal{M}$ the scaled
covariance converges, locally uniformly, to a covariance function of a
limiting stationary Gaussian field around $x$.

For the setup as above Nazarov-Sodin proved (\cite{So 2012}, Theorem $4$) that
\begin{equation*}
\E[ N(f_{L}) ] = \overline{c_{NS}}\cdot L^{m}+o(L^{m}),
\end{equation*}
for some $\overline{c_{NS}}\ge 0$ depending on the limiting fields only, namely their Nazarov-Sodin constants.
This result applies in particular to
random band-limited functions on a generic Riemannian manifold,
considered in ~\cite{SW}, with the constant $\overline{c_{NS}}>0$ strictly positive.

\section{Statement of results for arithmetic random waves}

\label{sec:torus}

Let $S$ be the set of all integers that admit a representation as a
sum of two integer squares and $n\in S$. The toral Laplace eigenfunctions
$f_{n}:\R^{2}/\Z^{2}\rightarrow \R$ of eigenvalue $-4\pi^{2}n$ may be expressed as
\begin{equation}
\label{eq:fm def}
f_{n}(x) = \sum\limits_{\substack{\|\lambda\|^{2}=n \\ \lambda\in\Z^{2}}}a_{\lambda}e^{2\pi i \langle x,\lambda\rangle}
\end{equation}
with some coefficients $a_{\lambda}$ satisfying
$a_{-\lambda}=\overline{a_{\lambda}}$.
We endow the space of
eigenfunctions with a Gaussian probability measure by making the
coefficient $a_{\lambda}$ i.i.d. standard Gaussian (save for the
relation $a_{-\lambda}=\overline{a_{\lambda}}$).

For this model it is known ~\cite{KKW 2013} that various local
properties of $f_{n}$, e.g., the total length of the
nodal line $f_{n}^{-1}(0)$, depend on the limiting angular
distribution of $\{\lambda\in \Z^{2}:\|\lambda\|^{2}=n\}.$ More
precisely, for $n\in S$ let
\begin{equation*}
\mu_{n} = \frac{1}{r_{2}(n)}\sum\limits_{\|\lambda\|^{2}=n}\delta_{\lambda/\sqrt{n}},
\end{equation*}
where $\delta_{x}$ is the Dirac delta at $x$, be a probability measure
on the unit circle $\mathcal{S}^{1}\subseteq\R^{2}$. Then in order to
exhibit an asymptotic law for the total length of $f_{n}^{-1}(0)$
such as its variance, or some other local properties of $f_{n}$, it is
natural  to pass to subsequences $\{n_{j}\}\subseteq S$ such that
$\mu_{n_{j}}$ weakly converges to $\mu$, a probability measure on
$\mathcal{S}^{1}$. In this situation we may identify
$\mu$ as the spectral density of the limiting field around each point of
the torus (when the unit circle is considered embedded $\mathcal{S}^{1}\subseteq\R^{2}$);
such a limiting probability measure $\mu$
necessarily lies in the set $\mathcal{P}_{Symm}$ of probability
measures on $\mathcal{S}^{1}$, invariant w.r.t.  $\pi/2$-rotation and
complex conjugation (i.e. $(x_{1},x_{2})\mapsto (x_{1},-x_{2})$).  In
fact, the family of weak-* partial limits of $\{\mu_{n}\}$
(``attainable'' measures) is known \cite{KuWi attainable} to be a
proper subset of $\mathcal{P}_{Symm}$.

Let $N(f_{n})$ as usual
denote the total number of nodal components of $f_{n}$.
An application of ~\cite{So 2012}, Theorem $4$ mentioned above
implies that if, as above, $\mu_{n_{j}}\Rightarrow \mu$ with $\mu$ some
probability measure on $\mathcal{S}^{1}$, we have
\begin{equation}
\label{eq:E[N(fm)]=cNSm+o(m)}
\E[N(f_{n})]= c_{NS}(\mu)n + o(n),
\end{equation}
with the same leading constant $c_{NS}(\mu)$ as for the
scale-invariant model \eqref{eq:EN(f;R)=cNSR^m+o(Rm)}.

\vspace{3mm}

In order to state our results first we will need the following notation:
let $$\nu_{0}=\frac{1}{4}\sum\limits_{k=0}^{3}\delta_{k\cdot\pi/2}$$
be the {\em Cilleruelo} measure \cite{C 1993}, and
$$\nu_{\pi/4}=\frac{1}{4}\sum\limits_{k=0}^{3}\delta_{\pi/4+k\cdot\pi/2}$$ be the {\em tilted} Cilleruelo
measure; these are the only measures in $\mathcal{P}_{Symm}$ supported
on precisely $4$ points.  We prove the following  concerning the
range of possible  constants $c_{NS}(\mu)$ appearing in
\eqref{eq:E[N(fm)]=cNSm+o(m)}.

\vspace{3mm}

\begin{theorem}
\label{thm:NS const torus}

For $\mu$ in the family of weak-* partial limits of $\{\mu_{n}\}$
  the functional $c_{NS}(\mu)$ attains an interval of the form
  $I_{NS}=[0,d_{max}]$ with some $d_{max}>0$. Equivalently,
\begin{equation*}
\E[ N(f_{n_{j}}) ] = c\cdot n_{j}+o(n_{j})
\end{equation*}
for some $\{n_{j}\}\subseteq S$, if and only if $c\in I_{NS}$.
Moreover, for $\mu\in\mathcal{P}_{Symm}$, $c_{NS}(\mu)=0$ if and only
if $\mu=\nu_{0}$ or $\mu = \nu_{\pi/4}$ (i.e., either the Cilleruelo
or tilted Cilleruelo measures.)

\end{theorem}

\vspace{3mm}

Theorem \ref{thm:NS const torus} is a particular case of a more general result
concerning arbitrary random fields on $\R^{2}$, presented in section \ref{sec:res scale inv}.
Concerning the maximal Nazarov-Sodin constant $d_{max}>0$, we believe
that the following is true.

\vspace{3mm}

\begin{conjecture}
\label{conj:max NS const symm}
For $\mu\in\mathcal{P}_{Symm}$, the maximal value $d_{\max}$ is uniquely
attained by $c_{NS}(\mu_{\mathcal{S}^{1}})$, where $\mu_{\mathcal{S}^{1}}$
is the uniform measure on $\mathcal{S}^{1}\subseteq\R^{2}$. In particular,
$$d_{\max}=c_{\text{RWM}}.$$
\end{conjecture}

\vspace{3mm}

\begin{question}
  What is the true asymptotic behaviour
  of $\E[f_{n_{j}}]$ for $\{n_{j}\}$ a Cilleruelo sequence, i.e.
  $\mu_{n_{j}}\Rightarrow \nu_{0}$?
  The latter might not admit an asymptotic law; in this case it would
  still be very interesting to know if the expected number of nodal
  components grows, in the sense that
  $$  \liminf\limits_{j\rightarrow\infty} \E[N(f_{n_{j}})]\rightarrow \infty. $$
  In fact, we have reasons to believe that the stronger bound
  $$ \E[ N(f_{n_{j}}) ]  \gg \sqrt{n_{j}} $$ holds.
\end{question}

\vspace{3mm}

Motivated by the fact that the nodal length variance only depends on the
first non-trivial Fourier coefficient of the measure ~\cite{KKW 2013}, and some other
local computations, we raise the following question.

\vspace{3mm}

\begin{question}
  Is it true that $c_{NS}(\mu)$ with $\mu\in\mathcal{P}_{Symm}$ supported on
  only depends on finitely many Fourier
  coefficients, e.g. $\widehat{\mu}(4)$ or
  $(\widehat{\mu}(4),\widehat{\mu}(8))$?
\end{question}

\vspace{3mm}

\section{Statement of results for random waves on $\R^{2}$}

\label{sec:res scale inv}

Let $\mathcal{P}_{R}$ be the collection of probability
measures on $\R^{2}$ supported on the radius-$R$ standard ball
$B(R)\subseteq \R^{2}$; by the scale invariance we may assume that
$R=1$, and denote $\mathcal{P}:=\mathcal{P}_{1}$.

\vspace{3mm}

\begin{theorem}
\label{thm:cNS cont}
The functional $$c_{NS}:\mathcal{P}\rightarrow\R_{\ge 0}$$ is
continuous w.r.t. the weak-* topology on $\mathcal{P}$.
\end{theorem}

\vspace{2mm}

Some aspects of the proof of Theorem \ref{thm:cNS cont}
can be found in section \ref{sec:proof cNS cont}.

\vspace{3mm}

\begin{proposition}
\label{prop:cNS(Cil)=0}
Let $\nu_{0}$ be the Cilleruelo measure on $\R^{2}$ as above. Then its Nazarov-Sodin
constant vanishes, i.e., $$c_{NS}(\nu_{0})=0.$$
\end{proposition}

Note that the result of Proposition \ref{prop:cNS(Cil)=0} is
in the same spirit as known constructions of (deterministic)
eigenfunctions of arbitrarily high energy with few or bounded number
of nodal components that arise in eigenspaces with spectral measure given
by the Cilleruelo measure (see the recent manuscript ~\cite{BeHe}).
Proposition \ref{prop:cNS(Cil)=0} can be proved by either considering
an explicit construction of a random field $f$ with the given spectral
measure $\nu_{0}$ and noting that for this model there are a.s. no
compact nodal components, or, alternatively, by a local
computation, e.g. of the number of ``flips", i.e. points $x$ with
$f(x)=\frac{\partial}{\partial x_{1}}f(x)=0$.

\vspace{3mm}

Combining Theorem \ref{thm:cNS cont}, Proposition
\ref{prop:cNS(Cil)=0}, and using the convexity of $\mathcal{P}$, we
obtain the following corollary.

\vspace{2mm}

\begin{corollary}
The Nazarov-Sodin constant $c_{NS}(\rho)$ for $\rho\in \mathcal{P}$
attains an interval of the form $[0,c_{\max}]$ for some
$0<c_{\max}<\infty$.
\end{corollary}

\vspace{3mm}

As for the maximal value of the Nazarov-Sodin constant, we make the
following conjecture.

\vspace{3mm}

\begin{conjecture}
\label{conj:max-ns-constant}
For $\rho\in\mathcal{P}$, the maximal value $c_{\max}$ is uniquely
attained by $c_{NS}(\rho)$ for $\rho$ the
uniform measure on $\mathcal{S}^{1}\subseteq\R^{2}$. In particular
(cf. Conjecture ~\ref{conj:max NS const symm}),
$$c_{\max}=d_{\max}=c_{\text{RWM}}.$$
\end{conjecture}

\subsection{On the proof of continuity}

\label{sec:proof cNS cont}

To prove Theorem \ref{thm:cNS cont} we follow the steps of
Nazarov-Sodin ~\cite{So 2012} closely, controlling the various error
terms encountered. One of the key aspects of our proof, different from
Nazarov-Sodin's, is proving a uniform version of
\eqref{eq:EN(f;R)=cNSR^m+o(Rm)} as below, perhaps of
independent interest.

\vspace{3mm}

\begin{proposition}
Let $f_{\rho}$ be a random field with spectral density $\rho\in\mathcal{P}$.
The limit
\begin{equation*}
c_{NS}(\rho) = \lim\limits_{R\rightarrow\infty}\frac{\E[N(f_{\rho};R)]}{R^{2}}
\end{equation*}
is uniform w.r.t. $\rho\in\mathcal{P}$. More precisely,
\begin{equation*}
\E[N(f_{\rho};R)] = c_{NS}(\rho)R^{2}+O(R)
\end{equation*}
with constant involved in the ``O"-notation universal.
\end{proposition}

\section{Acknowledgments}

The authors of the present manuscript would like to thank M. Sodin for
many stimulating and fruitful discussions, and insightful and critical
comments while conducting the research presented. We would also like
to thank Z. Rudnick for many fruitful discussions and his help in
improving the present manuscript, and P. Sarnak for his
support and interest in our work.

P.K. was partially supported by grants from the G\"oran
   Gustafsson Foundation, and the Swedish Research Council.
   The research leading to these results has received funding from the
European Research Council under the European Union's Seventh
Framework Programme (FP7/2007-2013) / ERC grant agreement
n$^{\text{o}}$ 335141 (I.W.). I.W. was
partially supported by the EPSRC grant under the First Grant scheme
(EP/J004529/1).

\end{document}